\documentstyle[preprint,aps,pre,psfig]{revtex}
\begin{document}
\title{Elastic constants from microscopic strain fluctuations} 

\author{Surajit Sengupta$^1$\thanks{{\it on leave from}: Material Science 
Division, Indira Gandhi Centre for Atomic Research, Kalpakkam 603102, India},
Peter Nielaba$^2$, Madan Rao$^3$\thanks{{\it on leave from}: Institute of 
Mathematical Sciences, Taramani, Chennai 600113, India} and K. Binder$^1$}

\address{$^1$ Institut f\"ur Physik, Johannes Gutenberg Universit\"at Mainz, 
55099, Mainz, Germany \\ $^2$ Universit\"at Konstanz, Fakult\"at f\"ur 
Physik, Fach M 691, 78457, Konstanz, Germany \\
$^3$ Raman Research Institute, C. V. Raman Avenue, Bangalore 560080, 
India}

\date{\today}

\maketitle

\begin{abstract}
Fluctuations of the instantaneous local Lagrangian strain 
$\epsilon_{ij}(\bf{r},t)$, measured with respect to a static ``reference''
lattice, are used to obtain accurate estimates of the elastic 
constants of model solids from atomistic computer simulations. The measured 
strains  are systematically coarse~- grained by averaging them within 
subsystems (of size $L_b$) of a system (of total size $L$) in the canonical 
ensemble. Using a simple finite size scaling theory we predict the behaviour 
of the fluctuations $<\epsilon_{ij}\epsilon_{kl}>$ as a function of $L_b/L$ 
and extract elastic constants of the system {\em in the thermodynamic limit} 
at nonzero temperature. Our method is simple to 
implement, efficient and general enough to be able to handle a wide class
of model systems including those with singular potentials without any 
essential  modification. We illustrate the technique by computing isothermal 
elastic constants of ``hard'' and ``soft'' disk triangular solids in 
two dimensions  from Monte Carlo and molecular dynamics simulations. We 
compare our results with those from earlier simulations and theory. 
\end{abstract}


\section{Introduction}
One is often interested in long length scale and long time scale phenomena 
in solids ( eg. late stage kinetics of solid state phase 
transformations\cite{abr,mart}; motion of domain walls interfaces\cite{domm}; 
fracture\cite{frac}; friction\cite{frik} etc.). Such phenomena are usually 
described by continuum theories. Microscopic simulations\cite{UMS} of finite 
systems, on the other hand, like molecular dynamics, lattice Boltzmann
or Monte Carlo, deal with microscopic variables like the positions and 
velocities of constituent particles and together with detailed knowledge of 
interatomic potentials, hope to build up a description of the 
macro system from a knowledge of these micro variables. 
How does one recover continuum physics from simulating the 
dynamics of $N$ particles? 
This requires a ``coarse-graining'' procedure in space (for equilibrium) 
or both space and time for non-equilibrium continuum theories. Over what coarse 
graining length and time scale does one recover results consistent with 
continuum theories? In this paper, we attempt to answer these questions for 
the simplest nontrivial case, namely, a crystalline solid, (without any point,
line or surface defects\cite{CL}) in equilibrium, at a  
non zero temperature far away from phase transitions. We show that  
coarse graining of microscopic local strain fluctuations obtained from 
configurations generated in a computer simulation, enables us to calculate 
elastic susceptibilities (compliances)  as a function of the coarse 
graining length. Detailed finite size scaling analysis of this data yields 
finally elastic constants of the solid -- the essential inputs to  
continuum elasticity theory\cite{LL}. The strain field (together with 
defect densities) constitute the coarse grained description of a 
solid essential for understanding solid state phase transitions like structural 
transitions\cite{STR,OT} and melting\cite{KTHNY,NH,GBM}. 

Calculation of elastic constants from simulations fall into two categories,
viz.  they are obtained either from,
thermal averages of fluctuations of the stress or the strain -- the 
so called ``fluctuation'' methods\cite{efluc},
or from the stress - strain curve as computed from a series of 
simulations\cite{estra}. 
Fluctuation methods, though requiring longer runs for accumulating 
statistically significant data, are often preferred because 
the entire matrix of elastic constants can be evaluated in a single run,
whereas in the latter method for every elastic constant an appropriate
strain (or stress) have to be applied. Also mapping out the stress - strain 
curve can be treacherous especially for soft systems where the possibility 
of setting up a plastic flow in the system is high\cite{bgw}. 
Though for most systems careful applications of either  
procedure should yield results of comparable accuracy, they suffer from 
some common limitations. Firstly, elastic constants are obtained for a 
particular system size $L$. In order to take the infinite size (thermodynamic)
limit one needs to simulate a sequence of systems with increasing values 
of $L$ -- often a computationally expensive proposition since equilibration of
large systems take increasingly larger times. Secondly, these procedures are 
not general and fail for model systems of particles interacting via 
singular potentials eg., the hard sphere\cite{FL,RC} or the hard disk\cite{WB} 
system where the 
instantaneous force on a particle is not well defined. To obtain elastic
constants of hard systems, one therefore needs to develop special 
methods\cite{FL,RC}. For instance, the elastic constants of the hard 
sphere system was calculated by Runge and Chester\cite{RC} by generalizing
a technique used previously for calculating the hard sphere pressure\cite{BH}. 
In this approach one tries to evaluate ensemble averages of quantities 
involving delta functions by calculating acceptance probabilities of virtual 
Monte Carlo steps. Such methods are cumbersome to use and may require ill 
defined averages of quantities whose variance has weak divergences\cite{RC}.
In contrast, the method presented here has several advantages. Firstly, 
elastic constants are obtained from a coarse graining procedure 
which automatically, gives the infinite system values. Secondly,
our procedure needs only the instantaneous particle configurations and
makes no reference to the potential or forces. It is therefore quite 
generally applicable to {\em any} system for which these configurations
or ``snapshots'' are known. Lastly, it will soon be evident that this 
technique is easy to use requiring a computational effort not much more than  
a calculation of, say, the pair distribution function\cite{UMS} for a given 
particle configuration. 

There are  three essential elements or steps in the method,
\begin{enumerate}
\item a procedure for calculating elastic strains from configurations.
\item the coarse graining procedure of averaging fluctuations of 
these strains over larger and larger sub~-blocks of length $L_b \leq L$  of a 
system of total size $L$ and calculating the values of these fluctuations 
in the thermodynamic limit
\item and, finally, converting the data for the strain fluctuations into 
elastic constants using  well known results of continuum elasticity 
theory\cite{LL,wal}.
\end{enumerate} 
We take up the second of these steps first as it is quite general and applicable
to fluctuations of any intensive variable. Indeed, variants of this method have 
been used extensively\cite{mcbin,blkl,blkc,rnb,span} in the past for obtaining 
finite size scaled susceptibilities and goes by the name of ``block~-
analysis''. In the next section (section II) we describe this block~-analysis 
technique for obtaining the 
susceptibility of the two~-dimensional Ising model\cite{rnb} for illustration. 
In section III we define the microscopic strains $\epsilon_{ij}({\bf r})$ and 
describe how we obtain them from our computer simulations. We then explain how 
to adopt the coarse graining scheme described in section II for strains to 
finally obtain the elastic constants. This is followed by our 
results for the hard disk and the inverse twelveth power ``soft'' disk 
systems in two dimensions. We conclude this paper with a discussion of 
these results and enumerating future directions.  

\section{Finite size scaling of fluctuations in subsystems}
The block analysis method has been used to obtain the compressibility of 
the Ising lattice gas\cite{blkl}, the two~-dimensional Lennard~-Jones 
fluid\cite{blkc,rnb} and fluids with internal classical and quantum degrees 
of freedom\cite{af,iqs}. Below, we describe a version of this method which 
allows us to explicitly and
systematically incorporate finite size effects arising both from a non~-zero
correlation length as well as special constraints due to the nature of the 
ensemble used. This treatment is analogous to the one used in Ref.\cite{span}
for analyzing density fluctuations in the two~-dimensional Lennard Jones 
fluid.

To begin, consider a general system described by a 
scalar order parameter $\phi(\bf r)$. We are interested in obtaining 
thermodynamic properties of this system in the disordered phase. It is 
therefore sufficient to use the following quadratic  
Helmholtz free energy functional, $F[\{\phi\}]$.   
\begin{equation}
\label{eq:landau}
F = k_B T\,\,\int d^dr \left( \frac{1}{2} r \phi^2 + \frac{1}{2} c \lbrace \nabla \phi(\bf{r}) \rbrace^2 \right)
\end{equation}
The correlation function $G_{\phi\phi}(q)$ implied by the above free energy 
in the high temperature phase ($<\phi> = 0$) is given\cite{CL} 
by the following Ornstein~-Zernicke form,
\begin{equation}
\label{eq:gq}
\beta G_{\phi\phi}(q) = <\phi_q \phi_{-q}> = {{\cal X}^\infty_{\phi\phi}}\frac{1}{1+(q\xi_{\phi\phi})^2}
\end{equation}
Where $\xi_{\phi\phi}$ is the correlation length ($=\sqrt{c/r}$), the 
(infinite system) susceptibility ${\cal X}^\infty_{\phi\phi} = < \phi^2>$ is, 
in turn, given by 
$\lim_{q\rightarrow0} \beta G_{\phi\phi}(q)$, $\beta = (k_B T)^{-1}$ and the 
angular brackets $<...> = \int [d\phi]\,\, \exp (-\beta F[\{\phi\}])$. 

Consider further that we 
want to measure this susceptibility from a computer simulation of an 
Ising model within a finite simulation box of size $L$. In the course
of the simulation we measure $\phi$ averaged within a 
sub-block of size $L_b \leq L$,
\begin{equation}
\bar{\phi} = L_b^{-d} \int^{L_b} d^dr \phi({\bf r}).
\end{equation}
Then the fluctuations of $\bar{\phi}$ measured within 
this block is,
\begin{eqnarray}
\label{eq:cor1}
<\bar{\phi^2}>_{L_b}{L_b}^{d} & = & {L_b}^{-d}\int^{L_b} d^dr'd^d r <\phi (r)\phi (r')>, \\ \nonumber 
	                & = & \int^{L_b}d^dr \beta G_{\phi\phi}(r), \\ \nonumber
	                & \equiv & {\cal X}^{L_b}_{\phi\phi} 
\end{eqnarray}
Where $G_{\phi\phi}(r)$ is the inverse Fourier transform of the correlation 
function defined in Eq.(\ref{eq:gq}) and is given\cite{CL}, quite generally, by,
\begin{equation}
\label{eq:cor2}
G_{\phi\phi}(r) =  \xi^{-2}{\cal X}^\infty_{\phi\phi} |r|^{(d-2)}Y(|r|/\xi_{\phi\phi}) 
\end{equation}
where,
\begin{equation}
\label{eq:cor3}
Y(\eta) = \int_0^{\infty} z^{d-1} dz \int (2 \pi)^{-d}d\Omega_d \frac{e^{iz \cos\theta}}{[z^2+\eta^2]}
\end{equation}

One expects therefore that as, $L_b \rightarrow L$ the block susceptibility 
${\cal X}^{L_b}_{\phi\phi} \rightarrow {\cal X}^\infty_{\phi\phi}$. The 
behaviour of the block susceptibilities, however, is strongly dependent on 
the ensemble\cite{endep} in which the simulation is carried out. For example,
using the lattice gas\cite{CL} language, in a ``grand canonical'' ensemble 
where the chemical composition $\phi$ of the lattice gas is allowed to
fluctuate  keeping the chemical potential difference fixed 
($= 0$ in this case) the block susceptibility does approach 
${\cal X}^\infty_{\phi\phi}$ for large $L_b$. In a canonical ensemble, however, 
the average of $\phi$ over the entire system is constrained to vanish for all
times. In such a case the behaviour of ${\cal X}^{L_b}_{\phi\phi}$ is 
more complicated but can, nevertheless, be explicitly determined
as follows.

Introducing the Lagrange multiplier $\kappa$ and defining the new correlator,
$$ G'(r)  = \int [d\phi] \phi(0) \phi(r) \exp (-\beta F - \kappa \int d^dr \,\,
\phi),$$
where the free energy $F$ is given by Eq.(\ref{eq:landau})), one shows that 
that $G'_{\phi\phi}(r) = G_{\phi\phi}(r)-\Delta_L$.  
The constant $\Delta_L$ is given by,
\begin{equation}
\Delta_L = \frac{1}{L^d} \int^{L} d^dr G_{\phi\phi}(r) 
\end{equation}
Repeating the analysis with $G_{\phi\phi}(r)$ replaced by $G'_{\phi\phi}(r)$ 
one obtains the desired finite size scaling form for
${\cal X}^{L_b}_{\phi\phi}$ in the canonical ensemble. We derive this form 
explicitly for the case of $d=2$ below.

The correlation function $G(r)$ is,
\begin{equation}
G(r) = \frac{2}{\pi}  \xi^{-2}{\cal X}^\infty K_0(r/\xi)
\end{equation}
where $K_0$ is a Bessel function and we have suppressed the subscripts in 
$G, {\cal X}$ and $\xi$ for simplicity. We have therefore,
\begin{equation}
\Delta_L   =  {\cal X}^\infty L^{-2} \Psi(L/\xi)
\end{equation}
with the function $\Psi (\alpha)$ being defined as,
\begin{equation}
\Psi(\alpha) = \frac{2}{\pi}\alpha^2 \int_0^1 \int_0^1 dx dy\,\, K_0 (\alpha \sqrt{x^2+y^2})
\end{equation}
As can be easily verified $\Psi(\alpha)$ goes to $1$ within a range 
${\cal O}(1)$. Using the above expressions we find finally,
\begin{equation}
\label{eq:main}
{\cal X}^{L_b} = {\cal X}^\infty [ \Psi(x L/\xi)-\Psi(L/\xi) x^2].
\end{equation}
This is the main result of this section. 

Note that the form given above implies ${\cal X}^{L_b} \rightarrow 0$ as $L_b 
\rightarrow L$. In general there will be higher order corrections to 
Eq.(\ref{eq:main}) coming from higher order correlations signifying 
the breakdown of the quadratic form for the free energy Eq.(\ref{eq:landau}).
These introduce terms of order $x^{(2N)}$ implying extra 
parameters (with the constraint ${\cal X}^{L} = 0$ being imposed to 
eliminate one of them). In Fig. (1) we have plotted ${\cal X}^{L_b}\times x$ 
against $x$ for various values of $L/\xi$. One observes that for $L/\xi \gg 1$
,$\Psi(\alpha) \rightarrow 1$ and 
Eq.(\ref{eq:main}) goes over to the following simple limiting form,
\begin{equation}
\label{eq:main2}
{\cal X}^{L_b} \times x = {\cal X}^\infty [x - x^3].
\end{equation}
One can then extract ${\cal X}^\infty$ from the slope of the linear region at 
small $x$. This is equivalent to the procedure followed 
for example in  Ref.\cite{rnb} for extracting the compressibility of the 
lattice gas at the critical density. Obviously, for $L/\xi \sim 1$, this
construction becomes less well defined and the finding the ``linear'' region
becomes subjective. Fitting the full data to the form given in 
Eq.(\ref{eq:main}) makes it possible to extract ${\cal X}^\infty$ even 
when the system size is not much bigger than the correlation length.
We illustrate this below using previously published data of Rovere, 
Nielaba and Binder\cite{rnb} obtained from simulations 
of the lattice gas model using spin exchange dynamics (canonical ensemble).

In Fig. (2) we plot ${\cal X}^{L_b}$ as a function of $L/L_b = 1/x$.
This choice of the axes is identical to the one used in the original 
work\cite{rnb} 
and is equivalent. The points are the data of Ref.\cite{rnb}
at various (dimensionless) temperatures $T/J \geq T_c/J = 2.2699$. The 
curves through the data are fits to Eq.(\ref{eq:main}). The extracted 
compressibility ${\cal X}^{\infty}$ are plotted in Fig. (3) and compared
with the previous estimates\cite{rnb} based on linear extrapolation and 
the theoretical curve\cite{thsus}. 
Though, for large $T$ where the correlation length $\xi$
remains small one gets identical results, ${\cal X}^{\infty}$ extracted from
the fits to the full curve continues to be closer to the theoretical (exact)
estimate as $T \rightarrow T_c$ (and consequently 
$\xi \rightarrow \infty$). Simultaneously, we also  obtain from 
Eq.(\ref{eq:main}) an estimate for the correlation length $\xi$ which is shown 
in Fig. (4). The 
agreement with theory\cite{thcor} is not as good. This is only to be expected
since the correlation length is  a more sensitive quantity 
than the susceptibility. Also, the fitting procedure cannot overcome 
errors built into the data because of critical slowing down near $T_c$
which is known to lead to a systematic underestimation due 
to the use of a biased estimator\cite{crsd}.

Further, close to $T_c$ the free energy functional
(Eq. (\ref{eq:landau})) ceases to be valid, the correlation 
function $G_{\phi\phi}(r) \sim |r|^{-(d-2+\eta)}$ and this affects our 
estimate for the correlation length. This is particularly important in
two dimensions\cite{thcor} where, in the Ising lattice gas, the 
exponent $\eta = 1/4$ 
indicates a rather large deviation of the correlation function from the
Ornstein Zernike behavior, Eq. (\ref{eq:gq}), while in $d=3$, $\eta \sim 0.03$
and hence Eqs.(\ref{eq:cor2}) and (\ref{eq:cor3}) should be more accurate.

Far away from a critical point, on the other hand,
none of these criticisms apply and accurate estimates for the susceptibility 
can be obtained even from relatively small systems even in $d=2$ using the 
ideas described here. 

\section{Strain fluctuations and elastic constants}
\subsection{Theory}
Imagine a system in the constant NVT (canonical) ensemble at a fixed density 
$\rho = N/V$ evolving in 
time $t$. For any ``snapshot'' of this system taken from this ensemble, 
the local instantaneous displacement field ${\bf u}_{\bf R}(t)$ defined 
over the set of lattice vectors $\lbrace {\bf R} \rbrace$ of a reference 
lattice (at the same density $\rho$) is
\begin{equation}
{\bf u}_{\bf R}(t) = {\bf R}(t) -  {\bf R} 
\end{equation}
where ${\bf R}(t)$ is the instantaneous position of the particle tagged 
by the reference lattice point ${\bf R}$. 
In this paper we concentrate only on perfect crystalline lattices;  
if topological defects such as dislocations are present the 
analysis below needs to be modified. The instantaneous Lagrangian 
strain tensor $\epsilon_{ij}$ defined at ${\bf R}$ is then given by\cite{CL,LL},
\begin{equation}
\epsilon_{ij} = \frac{1}{2} \left( \frac{\partial u_i}{\partial R_j} +
\frac{\partial u_j}{\partial R_i}+\frac{\partial u_i}{\partial R_k}
\frac{\partial u_k}{\partial R_j} \right)
\end{equation}
The strains considered here are always small and so we, hereafter, neglect the 
non~-linear terms in the definition given above for simplicity. The derivatives
are required at the reference lattice points ${\bf R}$ and can be calculated 
by any suitable finite difference scheme once ${\bf u}_{\bf R}(t)$ is known. 
We are now in a position to define coarse grained variables 
$\epsilon_{ij}^{L_b}$ which are simply averages of the strain over a 
sub-block of size $L_b$. The fluctuation of this variable then defines the 
size dependent compliance matrix $S_{ijkl} = <\epsilon_{ij}\epsilon_{kl}>$.
Before proceeding further, we introduce a compact Voigt notation 
(which replaces a pair of indices $ij$ with one $\alpha$) 
appropriate for two dimensional strains  - the only case considered in this 
paper. Using $1 \equiv x$ and $2 \equiv y$ we have,
\begin{eqnarray}
ij & = & \,\,\,\,11\,\,\,\,22\,\,\,\,12\,\,\,\, \\ \nonumber
\alpha & = & \,\,\,\,1\,\,\,\,\,\,\,\,\,2\,\,\,\,\,\,\,\,3 \nonumber
\end{eqnarray}
The only nonzero components of the compliance matrix are 
\begin{eqnarray}
\label{eq16}
S_{11} & = & <\epsilon_{xx}\epsilon_{xx}> \hbox{ $=  S_{22}$}\\ \nonumber
S_{12} & = & <\epsilon_{xx}\epsilon_{yy}> \hbox{ $=  S_{21}$}\\ \nonumber
S_{33} & = & 4<\epsilon_{xy}\epsilon_{xy}>  \nonumber
\end{eqnarray}
It is also useful to define the following linear combinations
\begin{eqnarray}
S_{++} & = & <\epsilon_{+}\epsilon_{+}>\hbox{ $=  2 (S_{11}+S_{12})$}\\ \nonumber
S_{--} & = & <\epsilon_{-}\epsilon_{-}>\hbox{ $=  2 (S_{11}-S_{12})$}\\ \nonumber
\end{eqnarray}
Where $\epsilon_{+} = \epsilon_{xx}+\epsilon_{yy}$ and
$\epsilon_{-} = \epsilon_{xx}-\epsilon_{yy}$.
Once the block averaged strains $\epsilon_{ij}^{L_b}$ are obtained, it is 
straight~-forward to calculate these fluctuations (for each value of $L_b$).

Since we are interested in the elastic properties of the system far away 
from any phase transition, a quadratic functional for the Helmholtz 
free energy $F$ suffices. We therefore use the following Landau 
functional appropriate for a two dimensional solid involving coarse grained 
strains to quadratic order in strains and its derivatives.
\begin{eqnarray}
\label{eq:laneps}
F & = & \int d^dr \,\,\, \lbrace c_1 \epsilon_+^2 + c_2 \epsilon_-^2 + c_3\epsilon_3^2 
+d_1 (\nabla \epsilon_+({\bf r}) )^2  \\
\nonumber
  &   & + d_2 (\nabla \epsilon_-({\bf r}) )^2 + d_3 (\nabla \epsilon_3({\bf r}) )^2 \rbrace
\end{eqnarray}
The coefficients $c_i$ and $d_i$ are, of course, not all independent because 
of further symmetries\cite{CL,LL} which are present for the two dimensional 
triangular solid though this does not influence the rest of our 
analysis. Note that,  Eq. (\ref{eq:laneps}) is simply a sum of three 
independent functionals in $\epsilon_{+}$, $\epsilon_{-}$ and $\epsilon_{3}$
each of the form given in Eq.(\ref{eq:landau}). Thus the analysis described
in detail in section II carries over almost unchanged. 
There is one small difference, however. Even in 
the canonical ensemble with fixed box dimensions, the 
microscopic strain fluctuations over the whole box are not zero but remains a 
small number of the order of $(a/L)^2$ where $a$ is the lattice parameter
so that,
\begin{equation}
\label{eq:constep}
\int^{L} d^dr <\epsilon_{\alpha}({\bf r}) \epsilon_{\beta}(0)> = C_{\alpha\beta} \left(\frac{a}{L}\right)^2. 
\end{equation}
Incorporating this modification into Eq.(\ref{eq:main}) we get,
\begin{eqnarray}
\label{eq:mainep}
S_{\gamma\gamma}^{L_b} & = & S_{\gamma\gamma}^\infty \left[ \Psi(x L/\xi)-\left(\Psi(L/\xi)-C\left(\frac{a}{L}\right)^2\right) x^2 \right] \\ \nonumber
                       &   & + {\cal O}(x^4).
\end{eqnarray}
Where the index $\gamma$ takes the values $+$,$-$ or $3$ and $x = L_b/L$ and 
we have suppressed subscripts on $\xi$ and $C$ for clarity.
The above equation Eq.(\ref{eq:mainep}) can now be used to obtain 
the {\em system size independent} quantities $S_{\alpha\beta}^\infty$,
$\xi$ and $C$. 

Once the finite size scaled compliances are obtained the elastic 
constants viz. the Bulk modulus $B = \rho \partial P/\partial \rho$ and 
the shear modulus $\mu$ are obtained simply using the formulae\cite{wal} 
\begin{eqnarray}
\label{eq:bul}
\beta B  & = & \frac{1}{2 S_{++}} \\
\label{eq:sh1}
\beta \mu & = & \frac{1}{2 S_{--}} - \beta P \\
\label{eq:sh2}
\beta \mu & = & \frac{1}{2 S_{33}} - \beta P
\end{eqnarray}
where we assume that the system is under an uniform hydrostatic pressure $P$.
The two expressions for $\mu$ should give identical results and constitutes 
an excellent internal check for numerical accuracy. 

This completes the description of our technique for obtaining elastic 
constants. It is obvious that throughout the derivation we do not ever refer
to the interparticle interactions. The only place where this information is
required is in the calculation of the pressure $P$ of the system --- a 
quantity routinely calculated in simulations. This, as we 
show later, is not a limitation even for systems with hard core potentials,
where the calculation of pressure is more involved\cite{RC}. 
We now describe our results for two model, two dimensional, solids. The 
elastic constants in each case are obtained for a high density perfect 
triangular solid. The generalization of this method for higher dimensions,
more complicated (less symmetrical) crystal types and even for amorphous 
materials is straightforward.

\subsection{Hard disks}
Hard disks interact with the extremely short ranged potential,
\begin{eqnarray*}
\beta V(r) & = & 0 \hbox{ for $r > \sigma$}\\
     & = & \infty \hbox{\,\,\,\,\,$r \leq \sigma$},  
\end{eqnarray*}
where $\sigma$ the hard sphere diameter can be used as the unit of length and
there is no energy scale. This makes the hard disk system particularly easy 
to simulate and is a popular testing ground for theories and 
techniques\cite{UMS}. On the other hand, a 
calculation of elastic constants in this system is difficult since a harmonic 
description for such a solid does not exist at zero temperature. The 
energy vanishes and the free energy is wholly entropic in 
origin. For this reason most computational methods for obtaining 
elastic constants which work for smooth potentials have to be either 
discarded or modified in order to study elastic 
properties of hard systems\cite{FL,RC,WB}. The only previous study of the 
hard disk 
elastic moduli is by Wojciechowski and Bra\'nkai\cite{WB}. These authors 
carried out a Monte Carlo simulation of $56$ hard disks in a box of variable 
shape in the constant stress ensemble\cite{UMS}. The elastic constants were 
obtained from the fluctuations in the shape of the entire box. Results were 
checked for finite size effects by repeating a few test runs for $24, 30$ 
and $90$ particles. 

In our method the results derived in the 
previous sections carry over without any change to systems of particles
interacting with hard potentials. Further we automatically calculate 
finite size scaled quantities. 
We present results for elastic constants of the hard disk system 
in two dimensions for a range of densities $0.95 < \rho^\ast < 1.1$
from Monte Carlo (MC) simulations with systems with sizes varying from 
$168$ to $12480$ particles. 
The simulations have been set up for a perfect triangular lattice in a 
slightly rectangular simulation box  with periodic boundary conditions. 
The number of cells along each side of the box is adjusted to make the 
simulation box as close to a square as possible. 
In the hard disk system one can considerably
accelerate the MC dynamics using special updating schemes\cite{UMS}.  
We use a square grid as an overlay on the simulation box, and choose 
the grid size to be small enough to accommodate only one 
disk. An occupancy list of the grid positions is computed and 
continuously updated. For an attempted move of a hard disk from one grid 
point to another, we first check if the new point is unoccupied and then 
check for overlap only with the particles occupying the neighboring grid 
points. At the density $\rho \sigma^2 = \rho^\ast$= 1 a standard simulation 
had the length of 
3$\times$ $10^6$ MC steps, every 10-th MC step has been taken for
averaging observables, in particular the block analysis was done
for $100$ random placements of the block.

In Fig. (5) we show the strain- strain fluctuations as a function
of the relative sub-block size computed in a system with $N=3120$ hard disks
at a density of $\rho^{\ast}=1.0$.
The data are fitted to Eq.(\ref{eq:mainep}) where we keep terms
up to order $(L_b/L)^4$.
The fits to Eq.(\ref{eq:mainep}) are excellent and the infinite system
compliances $S_{\alpha\beta}^\infty$ can be determined immediately. 
To determine the sensitivity of our results to the total system size 
we have also simulated systems with $N= 12480, 780$ and an extremely small 
system consisting of $N= 168$ particles. 
Using the ``infinite'' system values determined from the 
$N= 3120$ particle system it is, in principle,  possible to predict the 
behaviour of the strain fluctuations $S_{33}(L_b)$ for the other systems.  
In Fig. (6) we have plotted $S_{33}(L_b)$ for $12480, 3120, 780$ and $168$ 
particles. The points are the simulation data while the curves are fits to 
Eq.(\ref{eq:mainep}) where the value of $S_{33}^{\infty}$ was constrained to 
be fixed at that obtained from the data for $3120$ particles.
The results are seen to be almost insensitive to the total system size as 
expected.

Once the compliances $S_{\gamma\gamma}$ are obtained the elastic constants,
in units of $k_B T/\sigma^2$, can be derived immediately using 
Eq.(\ref{eq:bul}-\ref{eq:sh2}). Our results for 
densities other than $\rho^{\ast} = 1.0$ were obtained from  
Monte Carlo simulations for $N= 3120$ hard disks. To obtain the pressure,
which is required for evaluating the shear modulus $\mu$,
we simply integrate our bulk modulus $B$ (independent of the pressure 
in two dimensions)  starting from the rather high 
density $\rho^{\ast} = 1.1$ where the free volume\cite{FV} expression for 
the pressure $\beta P \sigma^2 = P^{\ast} = 2 \rho/(2/\sqrt{3}\rho -1)$ is 
accurate. Our results\cite{sco} 
for the equation of state for the hard disk system is shown in Fig. (7) and 
those for the elastic constants are shown in Fig. (8). The two expressions 
for the shear modulus in Eq.(\ref{eq:sh1},\ref{eq:sh2}) give almost identical results
and this gives us confidence about the internal consistency of our method.
We have also compared
our results to those of Wojciechowski and Bra\'nka\cite{WB}. We find that  
while their values of the pressure and bulk modulus are in good agreement 
with ours (and with free volume theory) they grossly overestimate the shear 
modulus. This is probably due to the extreme small size of their systems 
and/or insufficient averaging. Our results for the sub-block analysis shows 
that finite size effects are non~-trivial for elastic strain fluctuations and
they cannot be evaluated by varying the total size of the system from $24$
to $90$, an interval which is less than half of a decade\cite{smal}.
One immediate consequence of our results is that the Cauchy 
relation\cite{WB,BW} $\mu = B/2-P^{\ast}$ is seen to be valid upto $\pm 15 
\%$ over the entire density range we studied (see Fig. (9) ) though there is 
a systematic deviation which changes sign going from negative for small 
densities to  positive as the density is increased. This is in agreement 
with the usual situation in a variety of real systems\cite{BW} 
with central potentials and highly symmetric lattices and in disagreement 
with Ref.\cite{WB}. We have also compared our estimates for the 
elastic constants with the density functional theory (DFT) of 
Rhysov and Tareyeva\cite{rt}. We find that both the bulk and the 
shear moduli are grossly overestimated - sometimes by as much as 100\%. 

\subsection{Soft disks}
The system of particles in two dimensions interacting by a purely
repulsive inverse $12^{th}$ power pair potential $v(r_{ij})$ of the
form given by,
\begin{equation}
v(r) =  e \left( \frac{\sigma}{r} \right)^{12}
\end{equation}
has been studied\cite{bgw,r12mel} quite extensively. This system
has the advantage of being realistic without being too complicated,
since the form of the potential ensures that the entire equation
of state can be determined from that of a single isotherm\cite{bgw}. The
quantities $e$ and $\sigma$ sets the scales for energy and
distance respectively and can be both set equal to unity.
Both the zero and the finite temperature  elastic
constants of this system has been calculated over a large
range of densities\cite{bgw}. 
We have used this system to test the applicablity of our method to 
molecular dynamics simulations. Our results here are not as extensive
as in the hard disk case and we obtain elastic constants only for a single
state point. 
 
We simulate this system with a simple leap~-frog molecular
dynamics code incorporating a Nos\'e Hoover thermostat\cite{UMS} in order to
generate configurations in the canonical $NVT$ ensemble. The temperature
$T^\ast=k_B T/e$ is fixed at $1$ and the density $\rho^{\ast}=\rho \sigma^2$
at $1.05$ - a state sufficiently far from melting.
The number of particles were chosen to be $N = 780$ which is same as
that used in Ref.\cite{bgw} and corresponds to $26\times30$ unit
cells of a triangular lattice within a nearly square box. Starting from
the perfect lattice initial configuration, the system was equilibrated
for more than $10^6$ molecular dynamics time steps $\Delta t = 0.002$ (
in units of $\sqrt{m \sigma^2/e}$ where $m$ is the mass of the
particles). Subsequently, averages of thermodynamic quantities were
calculated over  $\approx 10^4$ uncorrelated configurations. Our results 
for the elastic compliances $S_{\gamma\gamma}$ are similar to that in 
the hard disk case and are shown in Fig. (10).

The final estimates for the bulk modulus $B = 77.96$ and shear modulus
$\mu = 23.34$ (in units of $k_B T/\sigma^2$) compare well  
with those of Ref. \cite{bgw} (viz. $B = 79.71$ and $\mu = 24.96$). 
Errors in our estimate for the bulk modulus  arise from statistical error 
in the acquired data and from the fits this is around $ 3 \%$. The 
shear modulus being a more sensitive quantity to compute is less accurate
and the two expressions for $\mu$ in Eq. (\ref{eq:sh1},\ref{eq:sh2}) now differ by 
$10 - 15 \%$ (the number quoted above is the average value). This maybe 
due to the smaller size of our system compared to the hard disk case, and
because subsequent configurations in molecular dynamics simulations 
are more correlated than in Monte Carlo.  

\section{Conclusion}
In summary, we show in this paper that a systematic coarse graining 
analysis of strain fluctuations yields elastic constants of solids 
from computer simulations to high accuracy. Our method incorporates
finite size scaling and produces elastic constants in the thermodynamic 
limit. The procedure is simple to implement and is general enough to 
be carried out for any system without modification.
Before we end this paper, the following comments are, perhaps, in order. 

\noindent
{\em The coarse graining variable:-~} Firstly, we
introduced this work as a ``test'' case of a general coarse graining procedure
which may be used to study the connection between microscopic computer 
simulations and long wavelength physics contained in continuum field theory 
approaches for phase transitions in solids. In this regard we would like to 
point out that we found, as usual,  the choice of the coarse graining 
variable to be important. Our 
results show that microscopic strains calculated by taking finite differences
of the displacement field constitute the correct variable to coarse grain over.
One could alternatively have averaged the displacement field ${\bf u}({\bf r})$ 
and calculated the strains from the coarse grained ${\bf u}$. This procedure 
happens to produce wrong results giving us elastic constants which are 
orders of magnitude too large. 

\noindent
{\em The strain correlation length:-~} 
One of the new results from our calculations is the correlation length
$\xi_{\alpha \beta}$ of strain fluctuations. This is found to be small -
$2~- 3$ lattice spacings -  for all components of the strain-strain correlation
functions and for both the systems. We have checked this independently
by explicitly measuring the correlation function $G_{33}(\| {\bf R}\|)$
defined for all the lattice vectors ${\bf R}$ of the two dimensional
triangular reference lattice in the soft sphere system. Though, the simple 
Ornstein Zernicke form
is inadequate to describe detailed features of the correlation
function and actual values for the correlation length is hard to estimate,
preliminary results from our simulations do support the above contention.
The correlation of the local density $\rho({\bf r})$ or its phase -- the 
displacement field ${\bf u}({\bf R})$ is, of course, long ranged, decaying
algebraically as it should in the solid state.

\noindent
{\em Renormalization by defects:-~} Our results for the elastic constants 
are obtained for high density perfect solids. In general a solid contains
point (vacancies and interstitials) and line (dislocations) defects. For 
example, in the hard disk case dislocations start appearing in our systems 
below a density of $\rho^{\ast} = 0.95$ a range we have not explored in this 
paper. In principle, there is no reason why our method cannot be adapted 
for systems containing defects though this involves considerable computational
complexity. There are basically two problems which arise when one wishes to
calculate elastic constants in the presence of defects. Firstly, we have to 
ensure that the density of each type of defect on the average attain their 
equilibrium value. This is nontrivial because nucleation barriers for defects 
densities are usually high which means large system sizes and long 
simulation times are required. Defect mobilities are sluggish in a solid 
but to ensure that they are fully equilibrated one has to wait long enough to 
allow a typical dislocation to travel a distance equal to the system 
size\cite{dmelt}. Secondly, once we are sure that our configurations contain,
on the average, the required defect densities, we have to evaluate the 
strain field in presence of these defects. This is, by far, the easier part. 
Defects can be either isolated point defects which arise 
in grand canonical simulations or vacancy~-interstitial pairs which 
nucleate  pairs of dislocations of opposite sign. The 
concentration of point defects in solids is vanishingly small $\sim 10^{-2} - 
10^{-4} $ atomic percent and they are not expected to change the elastic 
constants substantially. In any case they can be easily taken into account by 
redefining the reference lattice points. Topological defects introduce, in
addition, a singular part in the displacement field so that the strain 
field cannot now be evaluated simply by taking numerical derivatives. However, 
the singular contribution for each dislocation is known analytically - so that  
for a given configuration one can locate topological defects of each type and
treat the smooth and the singular parts separately. Lastly, we would like to 
point out that the above two problems viz. (1) obtaining an equilibrium defect 
concentration and (2) evaluating the finite size scaling of singular strain 
fields of configurations containing defects are present in {\em all} 
techniques for calculating elastic constants although they seem not to have 
recieved so far the attention they deserve. 

The renormalization of elastic constants 
by dislocations can also be obtained approximately by  using standard
recursion relations\cite{NH} once the core energy of a dislocation is
known (for instance from a separate simulation) and the present 
(manifestly ``bare'') values of the elastic constants are used as inputs.
Such a calculation has applications in a study of two dimensional 
melting\cite{KTHNY,NH}. We are, at present, carrying out detailed calculations
of the elastic constants, equation of state and dislocation core energies 
of the hard disk and inverse power triangular solids 
to investigate their melting behaviour. Further, near to the melting transition
additional finite size effects would be manifest (due to diverging correlation 
lengths) and they have to be taken into account separately which would 
necessarily involve simulations of a much larger scale than employed here.
For example, it is known that near the liquid to hexatic transition 
($\rho^{\ast} = 0.89$) $N=256^2$ hard disks are required\cite{jast} in order to 
reach the thermodynamic limit.  

\noindent
{\em Evaluation of local stresses:-~} Lastly, we would like to point out 
that our procedure can also be used in 
an ``inverse'' mode where knowing the elastic constants for a system, 
strain fluctuations can be 
used to calculate local stresses. This is especially useful in experimental 
studies\cite{maret,bech} of melting behaviour of colloidal particles where 
high quality digitized particle images are available. Efforts in this 
direction are in progress. 

\acknowledgements
We thank Hartmut L\"owen and M. M\"user for discussions. One
of us (S.S.) thanks the Alexander von Humboldt Foundation 
for a fellowship.
Partial support by the SFB~513 is gratefully acknowledged.


%
\begin{figure}
\begin{center}
\label{fig1}
\caption[]
{
A plot of ${\cal X}^{L_b}$ as a function of the relative sub-block
size $L_b/L$ as given by Eq.(\ref{eq:main}) for various values of the
parameter $L/\xi$ ($= 1,5,10$ \& $50$).
}
\end{center}
\end{figure}
\begin{figure}
\begin{center}
\label{fig2}
\caption[]
{
The susceptibility ${\cal X}^{L_b}(T)$ of a two dimensional  Ising system
in the constant magnetization ensemble as a function of the
relative sub-block size $L/L_b$ for values of the temperature $T/J > T_c/J$.
The symbols refer to simulation data of the critical lattice gas
simulation of Ref.~(23) and the lines are fits of Eq. (\ref{eq:main}) to these
data.
}
\end{center}
\end{figure}
\begin{figure}
\begin{center}
\label{fig3}
\caption[]
{
The estimates for $\chi_{\infty}(T)$ as obtained from the fits
shown in Fig. 2. ($\Diamond$) compared with the estimates of 
Ref.~(23) ($+$). Note that the finite size scaling analysis described 
here yields estimates which are closer to the exact result ( curve ).
}
\end{center}
\end{figure}
\begin{figure}
\begin{center}
\label{fig4}
\caption[]
{
The correlation length $\xi$ (in units of the lattice parameter $a$)
as a function of the scaled temperature 
$T/J$ of the lattice gas model at the critical density. Symbols: our 
estimates from fits to the data of Ref.~\cite{rnb} and curve: theory 
Ref.~\cite{thcor}.
}
\end{center}
\end{figure}
\begin{figure}
\begin{center}
\label{fig5}
\caption[]
{
Strain strain fluctuations or elements of the compliance matrix
$S_{\gamma \gamma}$ defined in Eq. (\ref{eq16}) ($\gamma = +,-$ or $3$) as a 
function of relative sub-block 
size $L_b/L$, symbols: simulation data; curve: fit to the scaling function
Eq. (\ref{eq:mainep}).
The results (symbols) shown are for a system of $N=3120$ hard disks at 
$\rho^{\ast} = 1.0$. 
}
\end{center}
\end{figure}
\begin{figure}
\begin{center}
\label{fig6}
\caption[]
{
The infinite system susceptibility $S_{33}^{\infty}$
obtained from the data for $N=3120$ particles is 
used to predict the finite size behaviour of $N=168, 780$ and $12480$ particles.
For $N = 168, 780$ and $3120$ the symbols are simulation data and the solid 
lines are the best fits to the form given in Eq. (\ref{eq:mainep})
where the parameter 
$S_{33}^{\infty}$ is kept fixed at the value obtained from fits to the data 
for $3120$ particles. 
For $N=12480$ we have acquired data (symbols) only for $L/L_b =$ an integer 
(between 4 and 35) and the dotted line is a straight line with the slope 
given by the same $S_{33}^{\infty}$.
}
\end{center}
\end{figure}

\begin{figure}
\begin{center}
\label{fig7}
\caption[]
{The equation of state of the hard disk solid, pressure $P^{\ast}$
as a function of the density $\rho^{\ast}$. We compare our results 
($\Diamond$) with those of Ref.\cite{WB} ($+$) and free volume 
theory (line).
}
\end{center}
\end{figure}

\begin{figure}
\begin{center}
\label{fig8}
\caption[]
{ The bulk ($B$) and shear ($\mu$) moduli in units of $k_B T/\sigma^2$ for 
the hard disk solid. Our 
results for $B$ ($\mu$) are given by $\Box$ ($\Diamond$). The values 
for the corresponding quantities from Ref.\cite{WB} are given by 
$+$ and $\times$. The line through the bulk modulus values is the 
analytical expression obtained from the free volume prediction for 
the pressure. The line through our shear modulus values is obtained from 
the free volume bulk modulus using the Cauchy relation $\mu = B/2 - P$. 
The error bars in our values for the shear modulus correspond to the two 
alternative formulae for evaluating $\mu$ as given in Eq.(\ref{eq:sh1},\ref{eq:sh2})
and represents our most conservative error estimates.
}
\end{center}
\end{figure}

\begin{figure}
\begin{center}
\label{fig9}
\caption[]
{ The percentage deviation $\Delta_c$ of our shear modulus values $\mu$ from
that obtained from our bulk moduli $B$ using the Cauchy relation $\mu = B/2 - P$as a funtion of the density $\rho^{\ast}$. 
The error bars in this graph correspond to the two
formulae for evaluating $\mu$ as given in Eq.(\ref{eq:sh1},\ref{eq:sh2}) as in 
the previous figure.
}
\end{center}
\end{figure}

\begin{figure}
\begin{center}
\label{fig10}
\caption[]
{
Same as in Fig. 5 but for $N=780$ soft disks interacting 
with the inverse $12^{th}$ power potential at
$\rho^{\ast} = 1.05$ and $T^{\ast} = 1$.
}
\end{center}
\end{figure}


\begin{references}
\bibitem{sco} The results presented in this paper for the perfect 
triangular solid in the hard disk system was obtained by imposing a 
``single cell occupancy'' constraint, rejecting all Monte Carlo moves which 
takes a particle out of its (hexagonal) Wigner~-Seitz cell. This ensures 
that no defects are produced during the Monte Carlo runs though the effect of
the constraint on elastic moduli for the densities  
considered is negligible. Below a density of $\rho^{\ast}= 0.95$ 
this constraint becomes increasingly relevant.
\bibitem{smal} To eliminate the effects of finite size completely one has 
 to use system sizes at least an order of magnitude larger than the correlation 
 length. Our fits to Eq.(20) yeilds correlation lengths of about $2 -3$ 
 lattice spacings, which implies  that we would need a system of at least 
 $N = (2x2x10)^2 = 1600$ particles taking care of periodic boundary conditions. 
 This is much larger than the sizes used by Wojciechowski and Branka.
\end{references}
\end{document}